# THE LINEAR-SIZE EVOLUTION OF
# CLASSICAL DOUBLE RADIO SOURCES

Mark J. Neeser

Max-Planck-Institut für Astronomie, Königstuhl 17, 69117 Heidelberg, Germany.
Electronic-mail: neeser@mpia-hd.mpg.de

Stephen A. Eales[1]

Department of Astronomy, University of Toronto, 60 St. George Street, Toronto, Canada,
M5S 1A7

J. Duncan Law-Green and J. Patrick Leahy

NRAL-Jodrell Bank, University of Manchester, Macclesfield, Cheshire SK11 9DL, UK

and

Steve Rawlings

Department of Astrophysics, University of Oxford, Nuclear and Astrophysics Laboratory,
Keble Road, Oxford, OX1 3RH, UK

astro-ph/9508036   8 Aug 95

---

[1]present address: Department of Physics and Astronomy, University of Wales College of
Cardiff, P.O. Box. 913, CF2 3YB, UK. Electronic-mail: sae@astro.cf.ac.uk

# ABSTRACT

Recent investigations of how the median size of extragalactic radio sources change with redshift have produced inconsistent results. Eales (1985) compared the radio and optical properties of a bright 3C and faint 6C sample and concluded that for a universe with $\Omega_0 = 0$ $D_{med} \propto (1+z)^{-1.1\pm0.5}$, with $D_{med}$ being the median size of the radio sources at a given epoch and $z$ the redshift. Oort, Katgert, and Windhorst (1987b), on the other hand, from a comparison of the properties of a number of radio samples, found much stronger evolution, with $D_{med} \propto (1+z)^{-3.3\pm0.5}$ for $\Omega_0 = 0$. In this paper we attempt to resolve the difference. We have repeated the analysis of Eales using the much improved data, in particular, the virtually complete redshift information that now exists for the 6C sample. Confining our analysis to FR2 sources, or "classical doubles", which we argue is the best-understood class of radio sources and the least likely to be affected by selection effects, we find $D_{med} \propto (1+z)^{-1.2\pm0.5}$ for $\Omega_0 = 0$ and $D_{med} \propto (1+z)^{-1.7\pm0.4}$ for $\Omega_0 = 1$. Moreover, in contrast to earlier studies, we find no intrinsic correlation between size and radio luminosity. We show that there is a selection effect which affects studies of this kind whose magnitude has not previously been realised, and has likely led to an overestimate in the strength of the size evolution found in previous investigations.

Our complete redshift information allows us to gain insight into our result by plotting a radio luminosity-size ($P$-$D$) diagram for the 6C sample, the first time this has been possible for a faint radio sample. The most obvious difference between the 3C and 6C $P$-$D$ diagrams ("the radio astronomers' H-R diagram") is the clump of sources in the 6C diagram at $D \sim 100$ kpc, $P_{151} \sim 10^{27} - 10^{28}$ W Hz$^{-1}$ sr$^{-1}$. These clump sources have similar sizes to the emission-line regions found around high-redshift radio galaxies, suggesting that the presence of dense line-emitting gas around high-redshift radio galaxies is responsible for the size evolution. We show that this explanation can quantitatively explain the observed size evolution, as long as there is either little X-ray emitting gas around these objects or, if there is, it is distributed in a similar way to the emission-line gas: highly anisotropic and inhomogeneous.

## 1. Introduction

The original interest in how the sizes of extragalactic radio sources change with cosmic epoch was concerned with the possibility that they might be 'standard rods' that could be used to measure the geometry of the universe (Miley 1968; Legg 1970). However, early work using samples of quasars (Legg 1970; Miley 1971; Wardle & Miley 1974) found evidence that the physical sizes of radio sources were generally smaller at earlier cosmic epochs. Since that time, the interest has been in the form and strength of this cosmic evolution with the hope that, since extragalactic radio sources are tenuous clouds of plasma that often lie well outside the host galaxies, this information might reveal something about the evolution of the intergalactic medium. Since several groups have recently used the ROSAT satellite to detect X-ray emission from high-redshift radio galaxies (Crawford and Fabian 1993; Spinrad, private communication), there is now also the possibility that a combination of X-ray observations, models for extragalactic radio sources, and an accurate determination of the form and strength of the linear-size evolution might lead to a consistent picture of how both radio sources and the galactic atmospheres surrounding them evolve (Subrahmanian & Swarup 1990; Gopal-Krishna & Wiita 1991). However, a number of problems and selection effects have stood in the way of a reliable determination of the form and magnitude of the linear-size evolution. We will describe these briefly, giving a few exemplary references in each case. A fuller review of the extensive literature in this area is given by Singal (1988).

(1) First there is the problem that the early radio surveys, such as 3C and 4C, which measured only peak fluxes, were susceptible to missing large sources because of their tendency to underestimate the flux of any source larger than the survey beam size. Omission of large sources would clearly have a serious effect on studies of the evolution of sizes. Recent surveys that produce integrated fluxes rather than peak fluxes, such as 6C (Hales, Baldwin & Warner 1993, and references therein), are essentially free of this problem. Although, in principle, a large source could still be omitted from a flux-limited sample of 6C sources if its surface brightness is, at all points, below the limiting surface brightness of the 6C survey, this is unlikely to be a problem for a sample, such as the one considered in this paper, in which the flux limit is substantially higher than the limiting surface brightness of the survey. It is worth remembering, however, that there is still only one bright radio source sample that should be relatively free of the problem of missing large sources (Laing, Riley & Longair 1983; Riley 1989).

(2) Intrinsic to any sample that is limited by magnitude or flux is the problem that luminosity and redshift will be correlated, so that any dependence of linear-size on redshift may actually arise from each parameter being separately correlated with luminosity. Stannard & Neal (1977) and Wardle & Potash (1977) first noticed that the apparent trend for quasars to have smaller radio-sizes at high redshifts could just be the result of the quasars with higher radio luminosities having smaller radio-sizes.

(3) There is the difficulty of finding optical identifications and redshifts for complete samples of radio sources. The early researchers in this field (e.g. Miley 1968, 1971; Wardle & Miley 1974) confined themselves to investigating the linear-size evolution of the quasars, since identifying and measuring redshifts for these sources in a radio sample is more easily done than for similarly distant galaxies. By the end of the 1970's the investigation was extended to all radio sources, primarily due to the fact that by this time most of the sources in the bright 3C sample had been identified and had had spectroscopic redshifts measured. However, because of the strong correlation between luminosity and redshift it was not possible to obtain a useful result from the 3C sample alone. The procedure that the investigators followed was to use the *physical* sizes of the sources in the 3C sample to predict the *angular* sizes that the sources in fainter samples should have. By comparing the predicted angular sizes with the observed angular sizes, which were relatively easy to measure, it was possible to estimate the strength and form of the linear-size evolution (Downes, Longair & Perryman 1981; Kapahi & Subrahmanya 1982; Allington-Smith 1984). However, this procedure is clearly inferior to comparing several samples, all of which have complete optical identifications and redshifts.

(4) Samples selected at high frequencies contain more compact sources (A fuller discussion of what is meant by a 'compact' source, as well as of the other possible source morphologies, is given below) than samples at lower frequencies. Thus an investigation of size evolution that compared the properties of a bright, low-frequency sample with a faint, high-frequency sample would necessarily find size evolution, even when there is actually none. An additional complication is that for a given luminosity, a source in a faint sample will be at a higher redshift than a source in a bright sample, and thus the difference between the frequency of emission and the frequency of observation will be even greater for the fainter source. Investigations of size evolution have to take this frequency-dependent selection effect into account. A fuller description of this effect is given by Allington-Smith (1984).

Since there are now some faint radio samples for which the fraction of sources that have optical identifications and/or redshifts is fairly large (Eales 1985c; Allington-Smith 1984; Dunlop et al. 1989), it might be expected that reasonable agreement about the form and strength of the size evolution should now be possible. However, this is not the case. For example, Eales (1985c) compared a faint sample of 6C sources with the bright 3C sample and reached the conclusion that if the functional form of the size evolution is $D_{med} \propto (1 + z)^{-n}$, where $D_{med}$ is the median size of the sources at a particular redshift, $n = 1.1 \pm 0.5$ for $\Omega_0 = 0$. Oort, Katgert, & Windhorst (1987b; henceforth OKW), on the other hand, compared the properties of a number of samples of radiogalaxies and determined that for this functional form, and with the same assumption about cosmology, $n = 3.3 \pm 0.5$. Singal (1988), Subrahmanian & Swarup (1990), and Kapahi (1989) have reached similar conclusions to OKW, namely that $n \simeq 3$, again by comparing the properties of samples of radiogalaxies. Lest it be thought that there is overwhelming evidence on the side of very strong evolution, it should be noted that researchers who have compared the predicted distributions of angular sizes in faint samples with the observed distributions, have all come down on the side of the much weaker evolution, $n \simeq 1.0 - 1.5$ (Kapahi & Subrahmanya 1982; Fielden et al. 1983; Allington-Smith 1984). Thus there is considerable disagreement about the strength of the evolution.

In this paper we attempt to resolve the disagreement. The paper is in two parts. In the first part we repeat the analysis of Eales (1985c) using the much better optical and radio data that now exists for the faint 6C sample (There are now, for example, spectroscopic redshifts for >95% of the sources). Our analysis is essentially the same as that of Eales except that we restrict ourselves to the FR2 class of sources. There have been a number of schemes for classifying the radio structures of extragalactic radio sources. The most venerable, and probably the most useful, is that of Fanaroff & Riley (1974), who divided double sources into ones where the distance between the radio peaks is over 50% of the total extent of the source (FR2s, or 'classical doubles' or 'edge-brightened' sources) and those in which this distance is less than 50% of the total extent of the source (FR1s, 'edge-darkened' sources). This classification almost certainly has physical meaning because, first, double sources do fall naturally into these two categories—an FR2 usually has its radio peaks right at the edge of the source, not just slightly over 50% of the way to the edge—and, secondly, FR2 sources have higher radio luminosities than FR1 sources. The FR classification does not, however, embrace all radio sources; there are also 'compact sources', so called for the historical reason that they were unresolved by the Cambridge 5-km telescope and are therefore less than an arcsecond in size (Jenkins, Pooley & Riley 1977; Laing 1981). This category probably contains a number of physically distinct types of radio sources. Many of the compact sources have flat radio spectra, in contrast to the FR1 and FR2 sources, and when these are observed with high angular resolution they usually have the core-halo or core-jet radio structures (Laing 1981a). There is also a class of compact sources with steep radio spectra (Peacock and Wall 1982). The physical nature of these, and in particular whether they are the progenitors of large FR2s, is uncertain (Pearson, Perley & Readhead 1985; Fanti et al. 1989; Dallasca et al. 1995). There are several reasons for restricting our study to FR2s. First, it is easy to define an unambiguous angular size: the distance between the radio peaks, or 'hotspots'. Second, as we will argue in §3, studies that include other types of sources are vulnerable to finding excessive evolution because of a combination of cosmological surface-brightness dimming and instrumental effects. Third, to the extent that we understand the physical operation of any class of extragalactic radio source, we understand the operation of FR2 sources best (Eales 1992 and references therein). Fourth, it makes sense to restrict the analysis to one physically distinct class of sources as otherwise, any apparent evolution may just be due to the proportions of sources in the different classes changing with redshift. Finally, there is the important point that most of the 3C and 6C samples are FR2 sources, and therefore little statistical power is lost by restricting the investigation in this way.

In the second part of the paper we consider the implications of our results. We first consider possible explanations for the difference between our results and those of OKW (§3). We show the reason for the disagreement does not lie in the different statistical techniques used, but that there are two selection effects that may have caused OKW to overestimate the strength of the linear-size evolution. We then use the radio luminosity-size diagram for the 6C sample, the so-called "$P$-$D$ diagram", to further investigate the nature and cause of the evolution (§4). We assume a Hubble constant of 50 km s$^{-1}$ Mpc$^{-1}$ and a zero cosmological constant ($\Lambda$) throughout this paper.

## 2. Reanalysis of the data for the 3C and 6C samples

Eales (1985c) investigated size evolution by applying the Spearman partial-rank statistic (Macklin 1982) to the sample of bright 3C sources of Laing, Riley & Longair (1983) and the sample of faint 6C sources of Eales (1985a,b,c). The 3C sample contains all the sources with flux density at 178 MHz ($S_{178}$) greater than 10 Jy with declination > 10° and galactic latitude > |10°| (Laing, Riley & Longair 1983). The 6C sample is a flux-limited sample with $S_{151}$ ~2 Jy (A precise description of the flux limits and the area of sky covered is given in Eales et al. 1995). Because of the way both samples were selected, fewer sources with large angular size should have been omitted than from other samples (Problem 1 in §1). The frequency-dependent selection effect (Problem 4) is unlikely to be important because the frequencies at which the samples were selected are similar—and indeed this is shown by the proportions of sources in the different structural classes being similar (Laing, Riley & Longair 1983; Eales 1985a). Furthermore, since two samples selected at different flux limits give a greater range of radio luminosity at a given redshift, and a greater range of redshift at a given radio luminosity, the difficulty of disentangling linear-size correlations with redshift from correlations with luminosity (Problem 2) is considerably less than it would be for a single sample. Eales' study did, however, have two problems. First, no attempt was made to investigate separately the size evolution of the different structural classes, with the attendant difficulties discussed in §1. At that time this was necessary because the radio maps of the 6C sources were often not sufficiently sensitive, or did not have sufficient angular resolution, to unambiguously assign a source to a structural class. Second, the partial-rank analysis, which is still necessary to separate correlations with radio luminosity from those with redshift, is only straightforward as long as each source has a redshift, luminosity, and size. Eales overcame the problem that many of the 6C sources did not have redshifts by using a Monte Carlo analysis to generate sets of redshifts for these sources (Eales 1985c). Although this procedure should be adequate, it did produce an unwanted complexity to the analysis.

Since Eales' work a large amount of new data has become available for the 6C sample. Naundorf et al. (1992), for example, presented better radio maps for about half the sources. More recently we have been making a concerted attempt to obtain better data for this sample, including better radio maps and, more importantly, optical and infrared images and redshifts. We now have spectroscopic redshifts for >95% of the sources, which makes this only the second sample, after 3C, with virtually complete redshift information. The new data will be presented in a number of papers: the new radio maps in Law-Green et al. (1994); infrared images in Eales et al. (1995); and the new redshifts in Rawlings et al. (1994). A complete reference list for all the observations of this sample is given in Eales et al. (1995). In this section we repeat Eales' analysis using the new data for the 6C sample. In §2.1 we discuss some of the details of the provenance of the data used in our analysis. The statistical method is briefly described in §2.2 and the results in §2.3.

## 2.1. The Data

The basic samples are described in Eales (1985a), Eales et al. (1995) and Laing, Riley & Longair (1983). Of the original 67 6C sources listed in Eales (1985a), five have been omitted. Better radio and optical observations have shown that three of these consist of two or more sources confused together, one has been omitted because the original 151-MHz flux given by Eales (1985) is clearly too high and its new flux lies outside the sample limits, and one is still in the sample but is so close to a bright star that, despite a number of attempts, we have been unable to get either an infrared image or to measure a redshift. Of the 62 sources remaining in the 6C sample we have redshifts for all but two (Rawlings et al. 1994). Fortunately, the remaining two sources are relatively bright galaxies, and we have estimated redshifts for these using the Hubble diagram for radio galaxies (Eales 1985c). Of the 173 sources in the 3C sample, all have redshifts. Complete lists of the redshifts of the 6C sample, both spectroscopic and estimated, are given in Eales et al. (1995) and Rawlings et al. (1994). Most of the redshifts for the 3C galaxies are listed in Spinrad et al. (1985).

Up-to-date lists of the redshifts for 3C sources are kept by H. Spinrad (Berkeley) and J. Riley (Cambridge).

The radio fluxes and spectral indices used for the 3C sources are those listed by Laing et al. (1983). The radio fluxes and spectral indices used for the 6C sources are those listed in Eales et al. (1994), and are slightly different from those listed by Eales (1985a) due to a recalibration of the 6C survey and of the flux scale used for the Cambridge One Mile Telescope.

The most uncertain step in our present analysis lies in assigning sources to structural classes. Fortunately, this is a problem that exists primarily for small sources, since sources of large angular size usually definitively fall into one or other of the FR classes. For small sources there are a number of problems. First, if a source is not much bigger than the beam of the radio telescope, only very limited structural information is available. Second, even when the map has excellent angular resolution, the difficulty in measuring the absolute position of a galaxy or a quasar to better than one arcsecond implies that there is often ambiguity about whether a radio peak is a core or a hotspot. This astrometric uncertainty between the optical/IR image and the radio map often makes it difficult to determine whether a source is a core-jet source or a small double. Furthermore, since knowing the location of the active nucleus is key to applying the FR classification criterion, applying the criterion to a source of this nature is problematical. With this in mind we used the following procedure to assign sources to structural classes. (A reference list for the radio maps of the 6C sources is given in Eales et al. (1994), a reference list for maps of the 3C sources can be obtained from JPL). For large sources we simply used the FR criterion. For small ($\simeq 1$ arcsec) sources we placed every source in the FR2 class unless (a) it was a double source but had one component much brighter than the other, suggesting a core-jet structure, (b) the source had a flat radio spectrum, which would be extremely unusual for an FR2 source, or (c) the source had a clear core-jet morphology. Of the 62 sources, seven fall in one of these categories, while four others are classified as FR1s and 51 as FR2s. Of the 173 3C sources, 131 are in the FR2 class. There are only a small number of sources where there is any uncertainty about their structural class and, in most of these cases, the independent assignment of two of us (SAE and JPL) almost always agreed.

For the double sources we defined angular size to be the distance between opposite hotspots. This is preferable to the definition used by Eales (1985c), who took the angular size as being the largest extent of the radio emission, since if there is some low-surface brightness emission outside the hotspots (though not enough to turn the source into an FR1) the value measured for the largest angular size will be too dependent on the telescope used for the observation and on the redshift of the source (§3).

## 2.2.   The Partial Rank Analysis

In recent studies of size evolution the method used to disentangle the relationships between luminosity $(P)$, redshift $(z)$ and linear size $(D)$ has generally been to divide the sources into bins of radio luminosity and then to look for a relationship between $D$ and $z$ within each bin (OKW; Singal 1988). This method, however, has a number of flaws. First, a method that employs binning is intrinsically inefficient; information about the ordering of $D$, $z$, and $P$ within each bin is necessarily thrown away. Second, because the luminosity range in an individual bin must be relatively broad in order to contain a sufficient number of sources for a useful statistical result, this creates the possible problem of a residual $D$-$P$ correlation within the bins producing a spurious $D$-$z$ correlation.

A method that is better, in principle, is to use the Spearman partial- rank statistic, $r_{ab,c}$ (Macklin 1982). This statistic was used by Eales (1985c) and will be adopted for this paper. The nonparametric statistic is a way of testing whether there is any evidence that the variables $a$ and $b$ are correlated if $c$ is kept constant. An intuitive way of looking at this is the following. The Spearman partial-rank statistic is a function of the standard Spearman rank correlation coefficients, $r_{ab}$, $r_{ac}$ and $r_{bc}$, which just measure the strength of the correlations between the pairs of variables. The partial- rank correlation coefficient, in essence, is a way of examining whether the correlation between $a$ and $b$ is significantly better than that which can be explained by the combined effect of the two other correlations: between $a$ and $c$ and between $b$ and $c$. The advantage of using this statistic over the binning method is simply that all the data is used in a statistically unambiguous way. The practical statistic used is $N_{ab,c}$, which is a straightforward function of $r_{ab,c}$ (Macklin 1982), and is distributed normally about zero with unit variance.

## 2.3.   Results

Table 1 shows our results for two geometries of the universe—$\Omega_0 = 1$ and $\Omega_0 = 0$. Given that our partial-rank analysis clearly indicates strong evidence for size evolution, we fitted the standard functional form, $D \propto (1 + z)^{-n}$, to the data in the following way. We multiplied the individual sizes of our sources by $(1 + z)^n$ and then adjusted $n$ until the value of $N_{Dz,P}$ produced by the partial-rank analysis was 0. We found the errors on $n$ by determining when $N_{Dz,P}$ became $\pm 1$ (Eales 1985c). The values of $n$ are also given in Table 1. Our values of $n$ are slightly higher than those found by Eales (1985c) and are much lower than those found in the recent work of Singal (1988) and OKW, which typically found $n \simeq 3$.

Singal (1988) claimed that whereas radio galaxies show strong size evolution, quasars show little or no size evolution. We cannot address this question directly because there are too few quasars in our samples, but we can examine the effect on the partial-rank analysis of removing the quasars. The results are shown in Table 1. Without the quasars the evolution is stronger, although still much less than the value found by OKW, giving some support to Singal's claim.

The results in Table 1 show no evidence for a correlation between radio luminosity and size that is not caused by the two variables being separately correlated with redshift. This is in contrast to the conclusion of OKW and Singal (1988). OKW found that, at constant redshift, the relationship between radio luminosity and size could be represented by the function $D_{med} \propto P^m$, with m = $0.3 \pm 0.05$. We fitted the same function to our data using a similar technique to that used for fitting linear-size to redshift. We multiplied each of our sizes by $P^{-m}$ and then adjusted $m$ until the value of $N_{DP,z}$ produced by the partial-rank analysis was 0. The errors on $m$ were found by determining when $N_{DP,z}$ became $\pm 1$. We found that $m = 0.06 \pm 0.09$ for $\Omega_0 = 1$ and $m = 0.04 \pm 0.09$ for $\Omega_0 = 0$, the latter being the cosmology used by OKW. Thus our results for both the strength of the linear-size evolution and the relationship between radio luminosity and size disagree with those of OKW.

## 3. Why are the estimates of the strength of the size evolution so different?

One possible explanation of the different results is that our study and that of OKW are examining the changes in $D_{med}$ over different ranges of redshift and radio luminosity. For example, the most distant source in the sample of OKW is at z $\simeq 0.75$, whereas the median redshift in the 6C sample is $\simeq 1$. The discrepancy between our result and that of OKW could then be explained if the relation between $D_{med}$ and $z$ is changing with redshift, and the relation between $D_{med}$ and $P$ is changing with radio luminosity. Although we cannot rule out this possibility, it would seem to require a somewhat unlikely conspiracy of parameters to produce the different results in both the $D$-$z$ and $D$-$P$ dependences.

A second possible explanation of the large difference between our estimate for the strength of the size evolution and that of OKW is that the difference may be a spurious one caused by the different statistical methods used. We checked this by applying the partial-rank analysis to the data of OKW. It is not trivial to reconstruct their data set, as the data is spread through many papers. We have estimated redshifts and angular sizes following methods as similar as possible to theirs, and although there are undoubtedly some differences between the dataset we have compiled and the original one, we are confident that the two are sufficiently close for our test to be meaningful. Table 2 shows the result of our partial-rank analysis. The value of $n$ we obtain, $n = 2.84^{+0.7}_{-0.6}$ for $\Omega_0 = 0.0$, compares well with the value obtained by OKW, and, like them, we also find evidence that there is an intrinsic correlation between luminosity and size. For consistency, we also checked the strength of the luminosity-size dependence by fitting the same functional form, $D_{med} \propto P^m$, as that used by OKW. We find that m $= 0.37 \pm 0.05$, again similar to the value obtained by OKW. The reason for our different estimates does not, therefore, lie in the different statistical approaches used. In the interest of thoroughness, we applied the values we obtained from the sample of OKW to our data by separately multiplying the linear sizes of 3C and 6C sources by $P^{-0.37}$ and $(1 + z)^{2.84}$. Our partial rank analysis rejects these values of n and m for our data at the 99.7% and 99.8% significance levels, respectively.

In a subject so fraught with selection effects, a third possibility is that either the size evolution has been overestimated by one group, or underestimated by another group, as a result of some unrecognised selection effect. We believe there are two effects that may have caused the evolution to be overestimated by OKW.

Suppose the sample that is being considered contains FR1 or core-halo sources. These are sources that, in contrast to FR2s, decrease in brightness with increasing distance from the active nucleus. Further suppose that there are two sources that are identical except that one is at a higher redshift. It is very easy to underestimate the angular size of the higher redshift source for two reasons, one fundamental and one instrumental. The fundamental reason is that surface brightness decreases with redshift as $(1 + z)^{-4}$. This means that if the radio observations of the two sources have the same surface-brightness sensitivity, the angular size measured for the source at high redshift will be smaller than it would be in the absence of 'cosmological surface brightness dimming': this means that if this effect is not taken into account (which it never has been) one will necessarily conclude that the linear-sizes of this type of source have experienced cosmological evolution. The instrumental effect is simply that sources in a faint sample, even when mapped with the VLA, rarely have maps with as high a surface-brightness sensitivity as sources in bright samples.

Figure 1 shows a plot of linear size verses redshift for the sources in the sample of OKW. There is a rather sudden drop in the linear size of sources at a redshift of z∼0.4, which is, interestingly, at the redshift that the sources in OKW's own sample begin to dominate. These are the faint sources drawn from the Leiden-Berkley Deep Survey (LBDS), which Oort et al. (1987) mapped with the VLA. We confirmed that it is this sample which is responsible for the strong evolution by repeating the partial-rank analysis excluding the LBDS sample. Without it there is virtually no evidence for any $D$-$z$ evolution at all. A further distinction between the LBDS sample and the other data sets used by OKW are the differences in median radio flux densities. The lower redshift subsamples of Gavazzi and Perola (1978) and Machalski and Condon (1985) have $S_{med}$(1.4 GHz)$\sim$1.8 x $10^4$ mJy and $S_{med}$(1.4 GHz)$\sim$6.7 x $10^2$ mJy, respectively, whereas the higher redshift LBDS subsample has $S_{med}$(1.4 GHz)$\sim$5.5 mJy. Therefore, beyond $z \sim$0.4 there is not only a strong fall-off in linear size, but also a decrease in median flux density of at least two orders of magnitude. Furthermore, because the median radio luminosity of this subsample is $P_{med} \sim$6.8 x $10^{23}$ WHz$^{-1}$sr$^{-1}$, it is the only one that lies below the Fanaroff-Riley break (Fanaroff and Riley 1974). As a result, a considerable fraction of the sources in the OKW subsample must be FRI-type objects. Figure 2, which shows the luminosity distributions for both the sources in OKW and the sources in the present study, confirms this: $\simeq$50% of the OKW sources lie below the FR break.

The OKW study should therefore be susceptible to the selection effects described above. Figure 3 shows how dramatic this effect can be. The VLA observations of the LBDS sources consisted of 30-minute integrations at 1.4-GHz, with the VLA in the A-configuration (Oort et al. 1987). The maps produced from these observations typically had an r.m.s. noise of 0.15 mJy beam$^{-1}$. In Fig. 3a we show a map made with the Cambridge One Mile Telescope (OMT) at 1.4 GHz of the typical FR1 source, 3C 31 (Burch 1979). Now imagine this source moved from its redshift of 0.0167 to z = 0.45, typical of the redshifts in the LBDS subsample, and observed using the telescope and procedure used for the LBDS observations. The physical resolution that the VLA observation will give at z = 0.45 will be exactly the same as the physical resolution that the OMT observation gives at the source's real redshift (not by coincidence). Because the physical resolution is the same, it is easy to estimate what the VLA map should look like, since this will just depend on the sensitivity of the observation. Fig. 3b shows our estimate of what this 3C 31 gedanken map should look like. Only the central part of the source is now visible, the remainder has too low a surface brightness. This is partially due to cosmological surface brightness dimming, but is mostly due to the VLA observation having poorer surface brightness sensitivity. If the reader is surprised that a map made with the geriatric Cambridge One Mile Telescope (c. 1962) should have better sensitivity than the VLA, note that the total integration time for the OMT observation was 96 hours. Oort et al. were aware of this problem and in many cases convolved their maps to a resolution of $\simeq$2-3.5 arcsec. This would have given a gain in sensitivity of approximately a factor of 2, which would have been enough to detect the southern peak in our gedanken source, but not the northern peak and all the extended emission. So, at the best, on the convolved map, the source would have appeared as two peaks of emission, thus being classified as an FR2, but with a much smaller angular size than the true extent of the source. This simulation shows that the effect of a combination of cosmological surface brightness dimming, limitations in telescope sensitivity, and source morphology can be severe. Although it is not enough to show that all the difference between our results and those of OKW can be explained in this way—for that we would have to carry out a simulation for every source in the 3C sample—it is enough to show that the strength of the evolution found by OKW must be an upper limit.

## 4. Discussion

A way of gaining insight into our result is to plot the radio luminosity-size ($P$-$D$) diagrams for the two samples. Baldwin (1982) first drew attention to the formal equivalence of the $P$-$D$ diagram to the H-R diagram, since, as in the H-R diagram, the two variables plotted are both observables and likely to be important parameters in any physical model of a source. In practice the $P$-$D$ diagram has not been as useful as the H-R diagram because, in contrast to the H-R diagram on which the very distinctive distribution of stars provides a strong constraint on the evolution of stars, the distribution of sources in the 3C $P$-$D$ diagram is much more homogeneous; there are no clearly empty regions or regions where there is a very high density of sources to provide strong constraints on the possible evolution of radio sources.

Figure 4 shows the $P$-$D$ diagrams for both the 3C and 6C samples, for both $\Omega_0 = 1$ and $\Omega_0 = 0$. The 3C and 6C diagrams look quite different. There are fewer very luminous sources and fewer sources with very low luminosities in the 6C diagram, and there appears to be a distinctive clump of sources at $D \sim 100$ kpc, $10^{27} < P_{151} < 10^{28}$ W Hz$^{-1}$ sr$^{-1}$ which is not seen in the 3C diagram. The explanation of the different luminosity distribution, which must be partially responsible for the appearance of a clump, is reasonably easy to understand. The space density of luminous radio sources increases much more strongly with redshift than that of low-luminosity sources (Dunlop & Peacock 1990). Since for a given radio luminosity the sources in a faint radio sample will be at a higher redshift than those in a bright radio sample, the faint radio sample will contain a smaller fraction of low-luminosity sources. The reason for the comparative absence of high-luminosity sources in the 6C sample is slightly more complicated. The space density of radio sources increases rapidly with redshift up to $z \sim 2$ (Dunlop & Peacock 1990). It is still an open question whether or not the space density starts to decline above this redshift (Dunlop & Peacock 1990; Rawlings et al. 1994), but in any event, any increase is much more gradual than at lower redshifts. The most luminous sources in 3C are already close to $z \sim 2$, whereas the sources with luminosities a factor of 10 lower are only at $z \sim 1$. In the 6C sample the most luminous radio sources will be at $z \sim 3$, whereas the sources with luminosities 10 times lower will be at $z \sim 2$. Since the space density of radio sources increases much more rapidly in the range $1 < z < 2$ than in the range $2 < z < 3$, the proportion of the most luminous radio sources should be less in the 6C sample. Although the striking appearance of the clump is partially caused by the lower proportions of low-luminosity and high-luminosity sources in the 6C sample, it cannot entirely be caused by this. The clump is too confined in both its $P$ and $D$ axes for this explanation to be the only origin for this feature. Figure 5 shows the distribution of linear size for the sources in 3C and 6C with radio luminosities in the range $7 \times 10^{26} < P_{151} < 7 \times 10^{27}$ W Hz$^{-1}$ sr$^{-1}$ ($\Omega_0 = 1$), the range of luminosity occupied by the clump. The distribution of size is much narrower for 6C than for 3C, showing that the clump is not simply caused by a different distribution in luminosity. The fact that the 6C sources have smaller sizes than the 3C sources of the same luminosity is, of course, the phenomenon detected by the partial-rank analysis.

Why then is there a trend for the sizes of classical doubles to decrease with redshift? The evolution of a radio source is thought to be governed by a combination of 'nature', the energy flowing out of the central massive object along the beam, and 'nurture', the pressure and density of the gas surrounding it (e.g. Scheuer 1974; Rawlings & Saunders 1991). We will consider, in turn, whether changes in these two factors could give rise to the size evolution.

In principle, size evolution could be produced if radio source beams are, on average, turned on for a shorter period of time at high redshift than at low redshift, thus restricting the maximum sizes attained by high-redshift classical doubles. Although this is an ad hoc explanation, and there is no reason to expect this to happen, we cannot rule it out. If we had firm expectations of what sources should look like after their beams have been turned off, and if we could predict the surface density of these 'dead sources' on the $P$-$D$ diagram, we could test this hypothesis, but as we show in the appendix our knowledge on this matter is too limited at present.

While there is no reason to expect 'nature' to change with redshift, there are direct observational clues that 'nurture' is changing. There are currently two pieces of evidence, one indirect and one direct, for this change in the environments of classical doubles. The indirect evidence is that the X-ray luminosity function for clusters shows strong negative evolution (Edge et al. 1990; Gioia et al. 1990) suggesting that, since luminous classical doubles tend to be preferentially found in clusters (Yates, Miller & Peacock 1989; Hill & Lilly 1990), high-redshift classical doubles will be surrounded by less dense gas than those at low redshift. The direct evidence is that high-redshift classical doubles are often surrounded by large, line-emitting nebulae (McCarthy, van Breugel & Kapahi 1991 and references therein). The latter is particularly interesting because the typical size of the emission-line nebulae is similar to the typical size of the clump sources in Figure 4. Furthermore, McCarthy et al. (1991) have found compelling evidence that this line-emitting gas is either influencing, or is tracing the gas that is influencing, the evolution of individual sources:

almost all high-redshift classical doubles have the hotspot closest to the nucleus on the side where the emission-line structure is brightest. The expansion of classical doubles is governed by the equation of ram-pressure equilibrium, $\rho v^2 = P$; $\rho$ being the density in the surrounding gas, $v$ being the expansion speed of the hotspot, and $P$ being the pressure in the hotspot. Therefore, this is exactly what one would expect: the hotspot on the side where the gas is densest, and thus the emission lines are brightest, will have travelled the shorter distance from the nucleus. Polarization studies have confirmed that the side on which the emission-line brightness is greatest is where the ionized gas is densest (Pedelty et al. 1989a,b).

If it is a change in the environment that is responsible for the linear size evolution, it is possible to estimate the size of the change that is needed in the following way. Let us just consider the clump sources, whose distributions of sizes are shown in Fig. 5. In the 3C sample these sources lie in the redshift range $0.4 < z < 1$; in the 6C sample these lie in the redshift range $1 < z < 2$. We will divide the sources into those whose sizes are less than 150 kpc and those whose sizes are greater than this, but less than 1 Mpc. We will take a very simple model for the density distribution around a source, and assume that the density within 75 kpc of a source is a constant, $\rho_-$, and that the density between 75 kpc and 500 kpc is another constant, $\rho_+$. From the equation of ram-pressure equilibrium, the time that a classical double will spend in a particular size range is $\propto \rho^{\frac{1}{2}}$, and thus the ratio of the time it will spend in the outer bin to the time it will spend in the inner bin is $\propto (\rho_+/\rho_-)^{\frac{1}{2}}$. The ratio of the numbers of sources in the two bins should also be proportional to this factor. In the 3C sample the ratio of the number of sources whose sizes are greater than 150 kpc, to the number that are less than this is 1.36; in the 6C sample this number is 0.30. Thus $(\rho_+/\rho_-)$ for the 3C sample is $\sim 21$ times that for the 6C sample, implying that the density distribution is a steeper function of distance from the galaxy for the high-redshift 6C sources than for the low-redshift 3C sources.

If the gas surrounding a high-redshift classical double were only the dense, $10^4$K gas that is responsible for the line emission, such a steepening in the density distribution would not be unexpected—there is usually no evidence for any line-emitting gas at distances more than 75 kpc from the nucleus. However, X-ray emission has been seen from high-redshift radio galaxies (Crawford & Fabian 1993; Spinrad, private communication) and, although it is often not possible to determine whether the X-rays are coming from the active nucleus or from a single hot gas phase, in the case of 3C 356 Crawford & Fabian (1993) have argued that the absence of absorption in the X-ray spectrum implies that the X-ray emission must be coming from hot gas. Furthermore, Fabian et al. (1987) have argued that the high-density, line-emitting gas must be confined by a lower density but much hotter gas, in order to stop the dense gas from dispersing on short timescales. We will consider what sort of two-phase model (line-emitting gas at $10^4$K and X-ray-emitting gas at $10^{7.5}$K) could produce the observed size evolution.

We will first consider a simple model in which a high-redshift classical double is surrounded by an X-ray-emitting halo similar to that observed around low-redshift classical doubles, but within the halo around the high-redshift source there are also lumps of line-emitting gas in pressure equilibrium with the hot gas. The luminosity in a collisionally-excited emission line, $L$, is given by

$$L \propto n^2 f V \tag{1}$$

(Osterbrock 1989), in which $n$ is the density, $f$ the filling factor, and $V$ the volume. The average density of the line-emitting gas, which is the important quantity for the ram-pressure equation is $fn$. Thus a measured line luminosity alone is not sufficient to obtain an estimate of $fn$. $f$ has been found directly for nebulae around low-redshift radio galaxies and lies between $10^{-6}$ and $10^{-4}$ (Heckman et al. 1982; van Breugel 1985). This is in reasonable agreement with more indirect estimates for high-redshift radio galaxies (Eales & Rawlings 1990; Chambers, Miley & van Breugel 1990). Assuming a value for $f$ of $10^{-5}$ and a characteristic line luminosity for a high redshift radio galaxy, a typical value for $n$ is $\sim 10^7$ $m^{-3}$. If this gas is being confined by the X-ray-emitting gas, the density of the X-ray-emitting gas will be $\sim 10^{3.5}$ $m^{-3}$, similar to the densities of the gas found around low-redshift classical doubles (Arnaud et al. 1984). However, because the filling factor for the X-ray-emitting gas must be close to unity, $fn$ is $10^{1.5}$ times greater for the X-ray-emitting gas than for the line-emitting gas, showing that in this model the line-emitting gas has a negligible dynamical effect on the source compared with the X-ray-emitting gas. Thus the mere addition at high redshift of some line-emitting gas to a non-evolving halo is not enough to explain the size evolution.

The X-ray-emitting gas must therefore also show strong evolution. One possibility is that the density distribution of X-ray gas is steeper at high redshift than at low redshift. However, although this would be enough to explain the size evolution, it would not explain the correlated asymmetries found by McCarthy et al. (1991) between the emission-line and radio structures; since the X-ray gas has a larger value of $fn$, the dynamical effect of the emission-line gas would be very small as compared to that of the X-ray-emitting gas. Something that would explain both results is if the X-ray gas around a high-redshift classical double is not distributed in a smooth halo like the gas around a low-redshift classical double, but instead as a series of clumps like the emission-line gas. There is some marginal observational support for this, as the one published X-ray map of a high-redshift radio galaxy does show several, although not very significant, peaks of emission (Crawford

& Fabian 1993). This still leaves the argument of Fabian et al. (1987) that these concentrations of gas should disperse on the sound-crossing timescale. It appears quite plausible to us, however, that these concentrations of gas are associated with peaks in the mass distribution, either galaxies or clouds of gas that are going to become galaxies, and that the ionized gas is confined by gravity. If the mass distribution around high-redshift classical doubles is not only lumpy, but also anisotropic, this would also provide a natural explanation of the alignments of the emission-line gas and the optical and infrared structures with the radio axis (Eales 1992).

This research was started while MJN and SAE were at the University of Toronto. SAE thanks the NSERC of Canada for a research grant and SAE and SR thank NATO for a collaborative research grant. DLG and SR were supported during this work by a PPARC research studentship and an Advanced Fellowship, respectively. SAE thanks Paul Alexander for a useful telephone conversation.

## A.   The Death of a Radio Source

If we knew how the appearance, luminosity and size of a classical double change with time once the beams are turned off, we could predict how the surface density of these 'dead sources' should vary over the $P$-$D$ diagram, and thus test the possibility that the linear-size evolution is caused by the beam on-time decreasing with redshift. When the beams are turned off, a classical double should pass through three evolutionary phases: (A) the hotspots, which are at substantially higher pressure than the radio lobes and the surrounding gas, will expand at the sound speed. Since for a relativistic plasma the sound speed is $c/\sqrt{3}$, the expansion timescale will be much less than the age of the source. The effect of adiabatic expansion on the relativistic plasma and the magnetic fields will cause the radio luminosity to decrease very rapidly but, since at low frequency the luminosity of the hotspots is only a small fraction of the total radio luminosity, the total radio luminosity of the source will not change very much. (This statement is true at least for the archetypical and closest luminous classical double, Cygnus A, but it is unclear at the moment whether it is also true for high-redshift classical doubles—Leahy, Muxlow & Stephens 1989). (B) Starting at the same time as the first phase, but taking much longer, the radio lobes will expand. For a source in which the energy density ($u$) in particles is equal to that in the magnetic field, the radio luminosity per unit volume is $\propto u^{\frac{7}{4}}$. For an expanding cloud of plasma and fields $P \propto D^{-4}$, and thus the radio source will follow a steep evolutionary track across the $P$-$D$ diagram. (C) The pressure in the lobes will also fall as $D^{-4}$ and eventually, when the pressure has declined to the level of the thermal pressure in the surrounding gas, the expansion will stop. After this, the size will stay the same and the luminosity will only slowly decrease due to radiative losses and the scattering of microwave-background photons by the relativistic electrons in the source. The evolutionary track across the $P$-$D$ diagram is shown in Fig. 4c. If there were any of these "post-main-sequence" sources in the predicted region of the $P$-$D$ diagram, this would be strong evidence for the hypothesis that the linear-size evolution is being caused by the beam on-time decreasing with redshift. There are no sources with the characteristics one would expect—radio lobes but no hotspots. However, it is unclear whether one would expect to see any because of Malmquist bias. For a flux-limited sample the volume of space in which a source can be seen is a strong function of luminosity (in a Euclidean universe, $\propto P^{\frac{3}{2}}$), and so if the luminosity of a source falls fast enough after the beams are turned off, sources like this should be very hard to find. The answer to this hinges on how long phase B lasts. In this phase the luminosity falls very quickly, whereas once a source is in phase C its luminosity will fall very slowly. The length of phase B depends on only one factor: how far the pressure in the lobes of a typical classical double is above the pressure in the surrounding gas. There is, of course, no consensus about this. Arnaud et al. (1984) found that the pressure in the lobes of the archetype classical double Cygnus A is only slightly different from the pressure in the surrounding gas, but as their lobes pressures were estimated from the minimum-energy argument, they strictly placed only a lower limit on the difference. In contrast, Miller et al. (1985) concluded that the lobe pressures in classical doubles are much higher than the pressure in the surrounding gas. Another argument that the thermal pressure in the surrounding gas is much less than the lobe pressure is that the axial ratio of a classical double does not appear to depend on its size (Alexander, priv. comm.). If the lobes of classical doubles are confined by thermal pressure, small young sources should be much rounder than large old sources. In summary, despite over two decades of observational and theoretical work on the physics of classical doubles, we still have a very limited idea of the evolution of a classical double once the beams turn off, and we cannot rule out this explanation of the size evolution.

Table 1: Results of Partial-Rank Analysis

| Source Type ($\Omega_0$) | $r_{Dz,P}$ | $N_{Dz,P}$ (prob) | $r_{DP,z}$ | $N_{DP,z}$ (prob) | Evolution Strength: n |
|---|---|---|---|---|---|
| Galaxies & Quasars (1) | -0.254 | -3.45 ($\simeq 0.05\%$) | 0.046 | 0.61 ($\simeq 54\%$) | $1.71^{+0.40}_{-0.48}$ |
| Galaxies & Quasars (0) | -0.159 | -2.12 ($\simeq 3.6\%$) | 0.027 | 0.36 ($\simeq 72\%$) | $1.22^{+0.48}_{-0.58}$ |
| Galaxies (1) | -0.286 | -3.57 ($\simeq 0.05\%$) | 0.101 | 1.23 ($\simeq 22\%$) | $1.96^{+0.43}_{-0.49}$ |
| Galaxies (0) | -0.207 | -2.54 ($\simeq 1\%$) | 0.090 | 1.12 ($\simeq 26\%$) | $1.57^{+0.50}_{-0.57}$ |

Table 2: Results of Partial-Rank Analysis of OKW Data

| Source Type ($\Omega_0$) | $r_{Dz,P}$ | $N_{Dz,P}$ (prob) | $r_{DP,z}$ | $N_{DP,z}$ (prob) | Evolution Strength: n |
|---|---|---|---|---|---|
| Galaxies (0) | -0.316 | -4.05 ($\simeq 0.006\%$) | 0.587 | 8.33 ($< 10^{-9}\%$) | $2.84^{+0.69}_{-0.64}$ |

Note. — $r_{ab,c}$ is the partial rank correlation coefficient for the triplet of variables, $a$, $b$ and $c$, and $N_{ab,c}$ is the function of the partial-rank correlation coefficient that is distributed like a normal distribution with a variance of unity (Macklin 1982). The probability of the value of $N_{ab,c}$ occurring by chance if there is no correlation between a and b, other than that caused by a and b each being separately correlated with c, is given in brackets.

# FIGURE CAPTIONS

**Figure 1:** Linear size verses redshift for the OKW sample. The different symbols show sources from the different subsamples used by OKW.

**Figure 2:** Histograms of radio luminosity at 1.4 GHz for the FR2 sources from our 3C and 6C (cross-hatched) samples (continuous line) and all the sources from OKW (dot-dashed line). The vertical dotted line shows the approximate dividing line between sources with the FR1 structure and those with the FR2 structure (Fanaroff & Riley 1974).

**Figure 3:** Simulation of the selection effect described in §3. Fig. 3(a) shows a 1.4-GHz map of the typical FR1 source 3C 31 (Burch 1979) made with the Cambridge One Mile Telescope. Fig. 3(b) shows our estimate of what this object should look like if it were moved to a redshift of 0.45 and observed with the VLA, using the same procedure as that used by Oort et al. (1987a) for the LBDS sources. Oort et al. (1987a) used the A-configuration of the VLA and integrated for typically 30 minutes. The r.m.s. noise on their maps was typically 0.15 mJy beam$^{-1}$. We have assumed $\Omega_0 = 1$. The lowest contour plotted in Fig. 3(b) is $2\sigma$.

**Figure 4:** Radio luminosity at 151 MHz verses physical size for both the 3C sample and the 6C sample, and for both $\Omega_0 = 0$ and $\Omega_0 = 1$ cosmologies. The filled circles show the FR2 sources, while the open circles show sources with other radio structures. The horizontal lines give an idea of the redshifts of the sources. These lines show the radio luminosities at which a source just bright enough to be included in the sample would be at redshifts of 0.5, 1, and 2. The dashed line in Fig. 4c shows the evolutionary track described in the appendix. The point at which the fast phase, B, turns into the slow phase, C, depends on the ratio of the pressure in the radio lobes to the pressure in the surrounding gas.

**Figure 5:** Histogram of size for the FR2 radio sources with luminosities in the range $7 \times 10^{26} < P_{151} < 7 \times 10^{27}$ W Hz$^{-1}$ sr$^{-1}$. The value of $\Omega_0$ assumed is one. The first figure is for the 3C sources, the second figure is for the 6C sources.

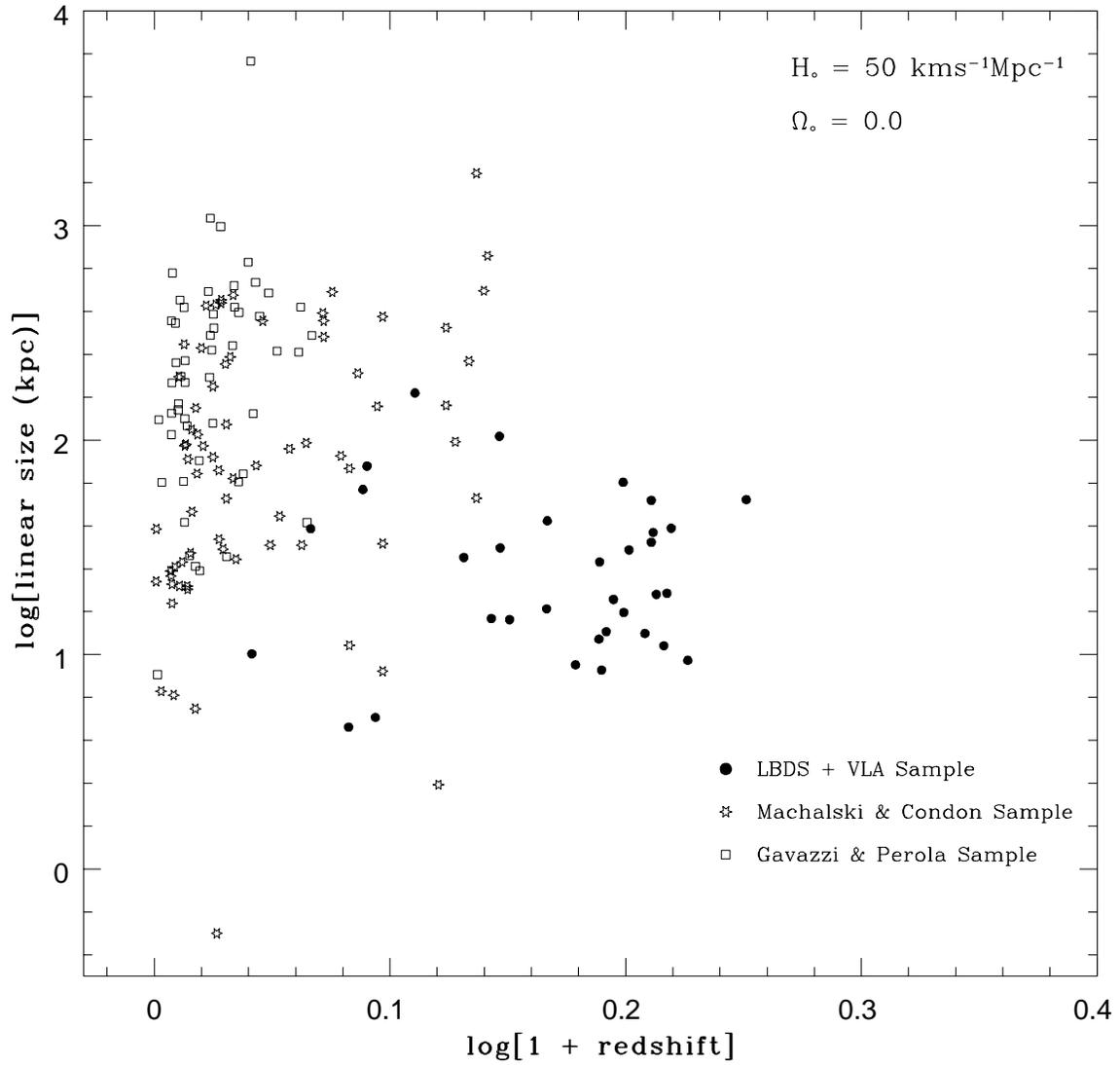

Figure 1

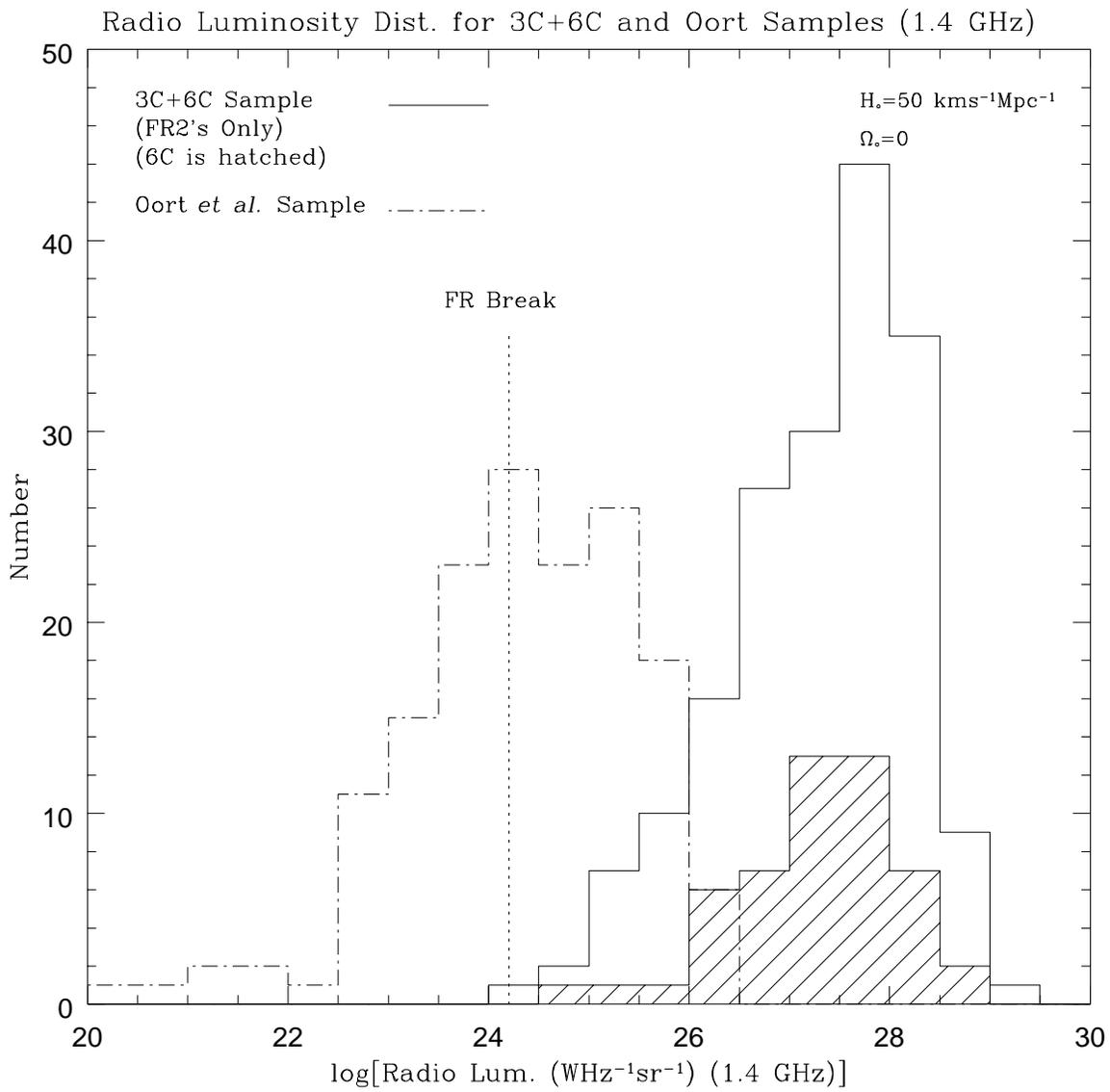

Figure 2

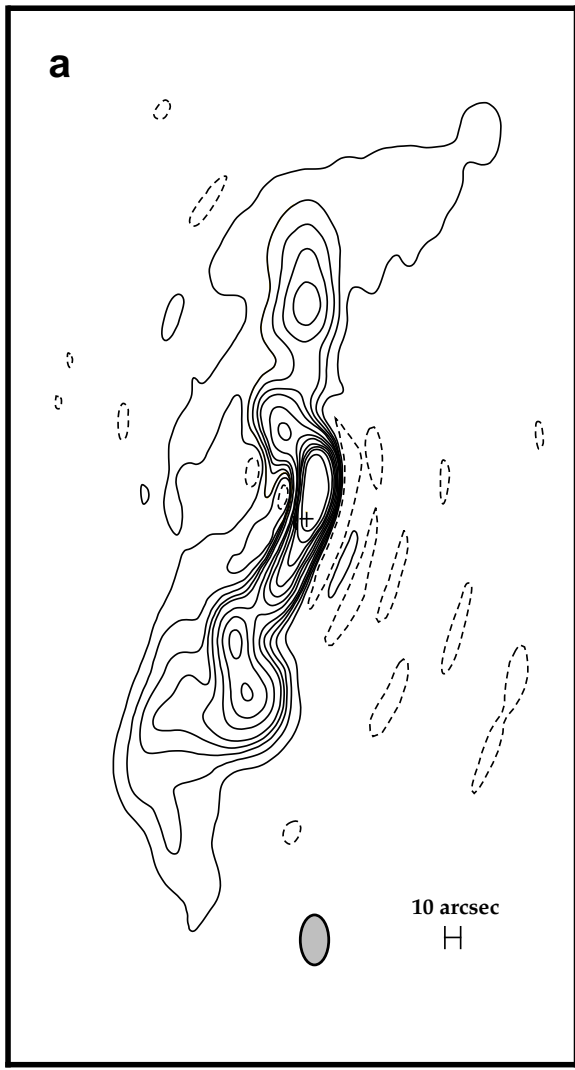

3C 31       1407 MHz

a

10 arcsec

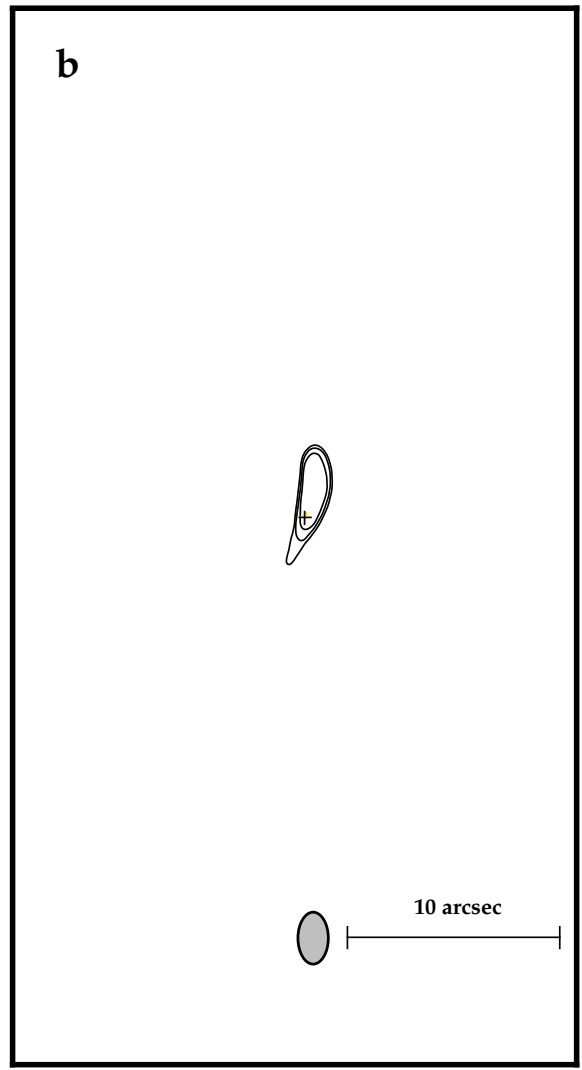

b

10 arcsec

Figure 3

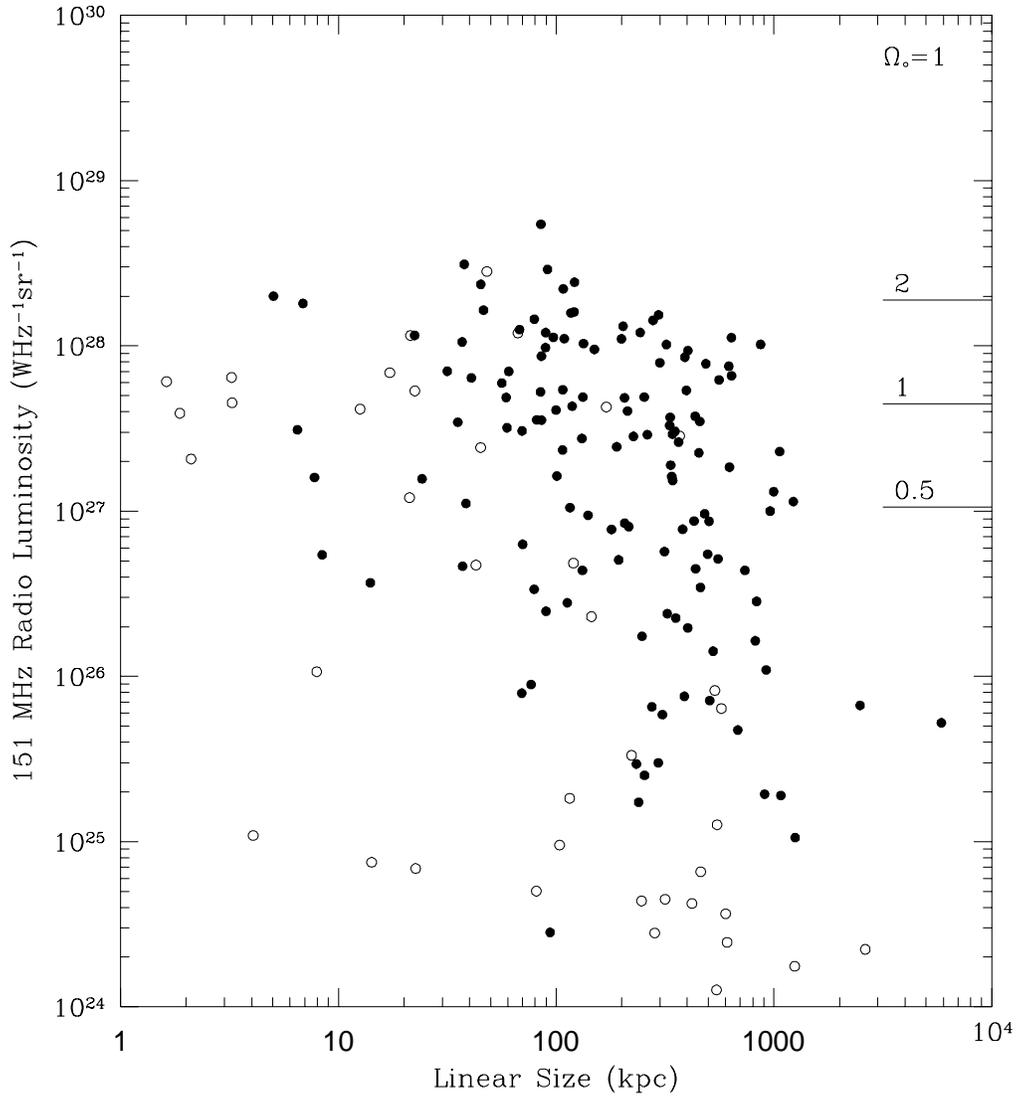

Figure 4a

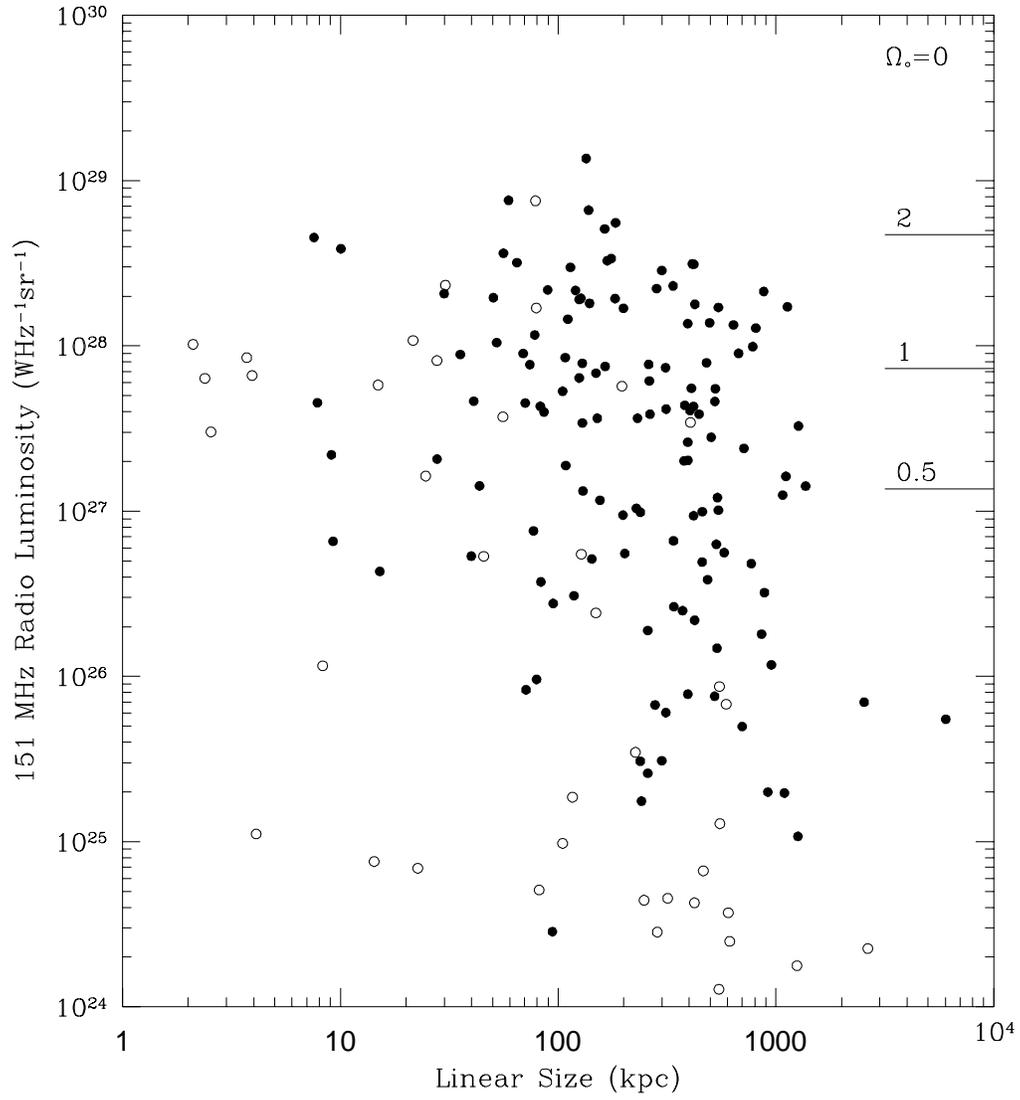

Figure 4b

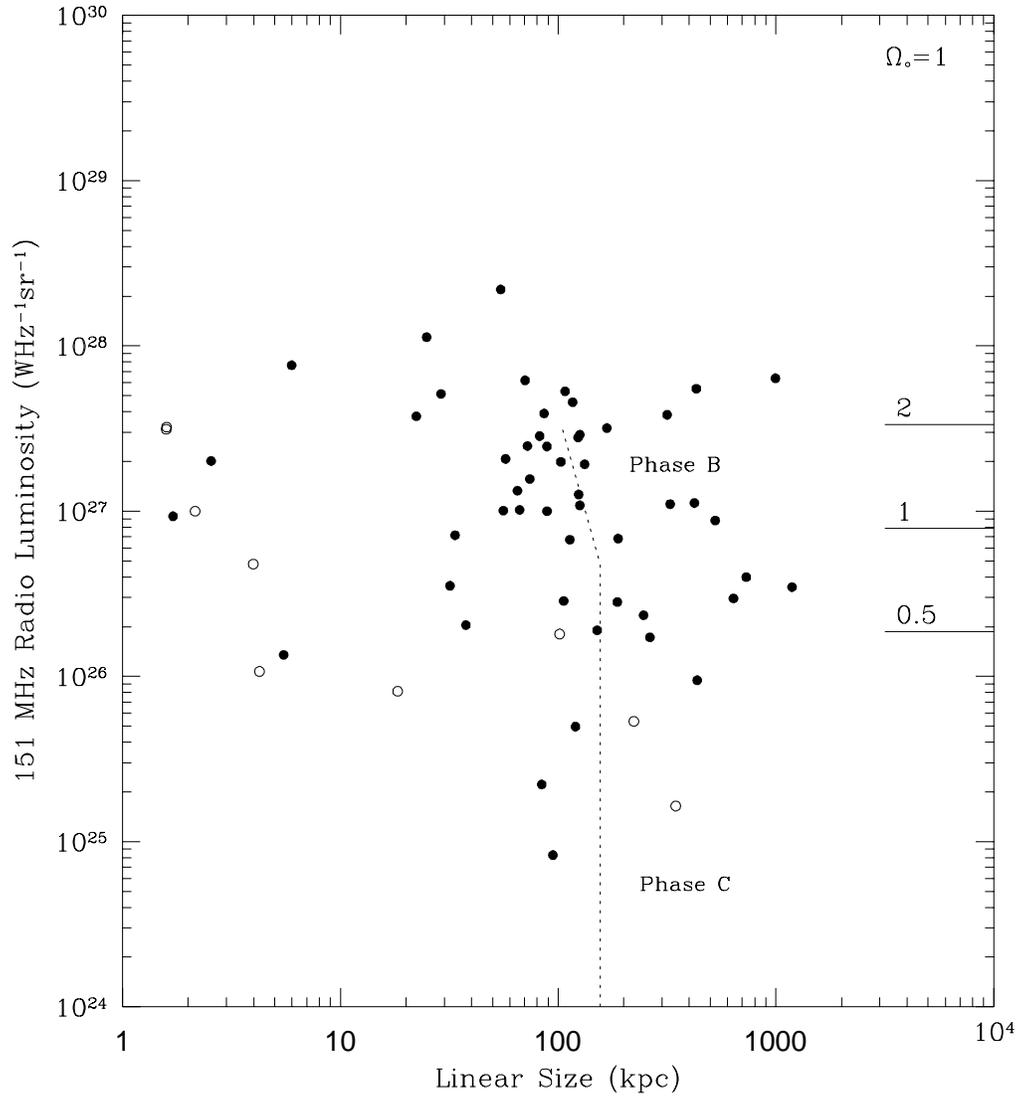

Figure 4c

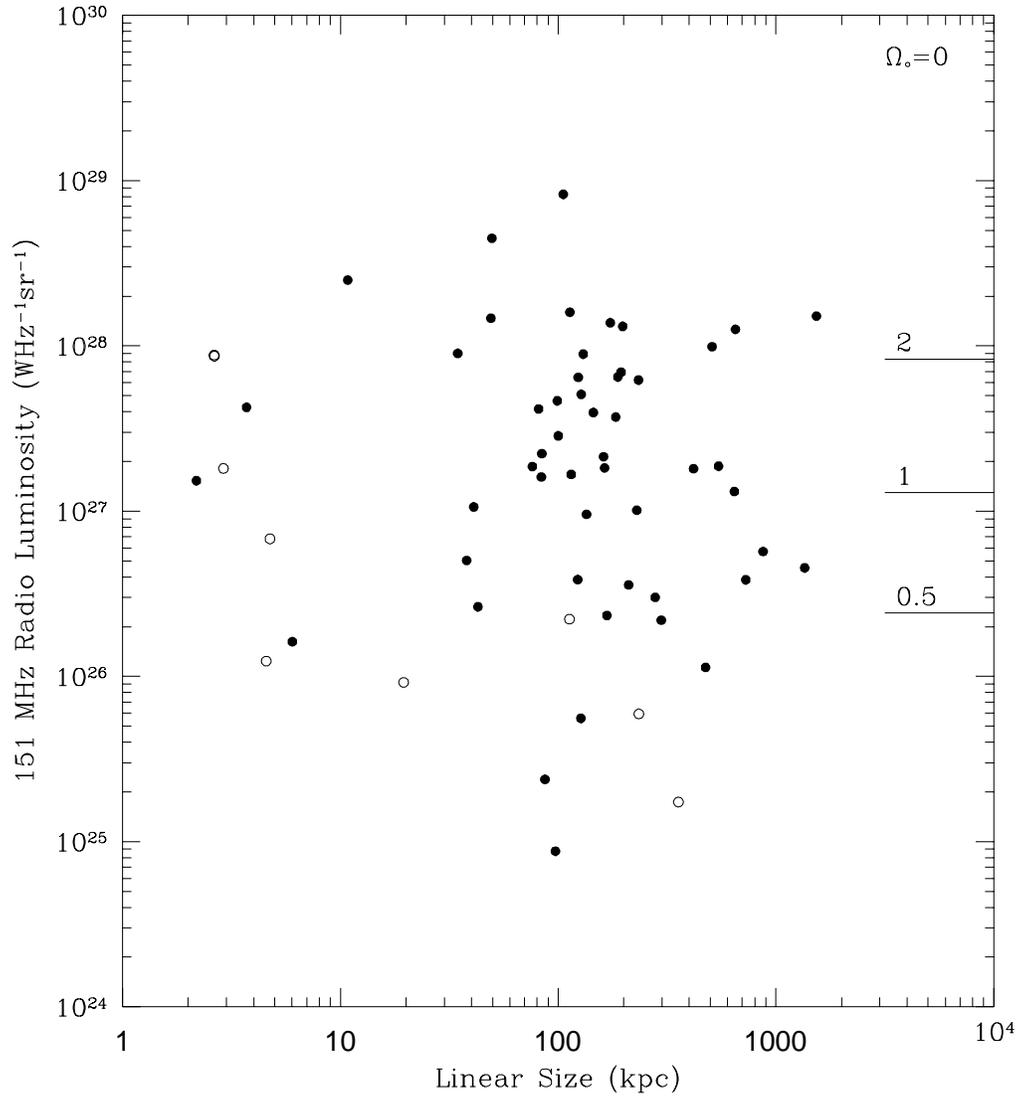

Figure 4d

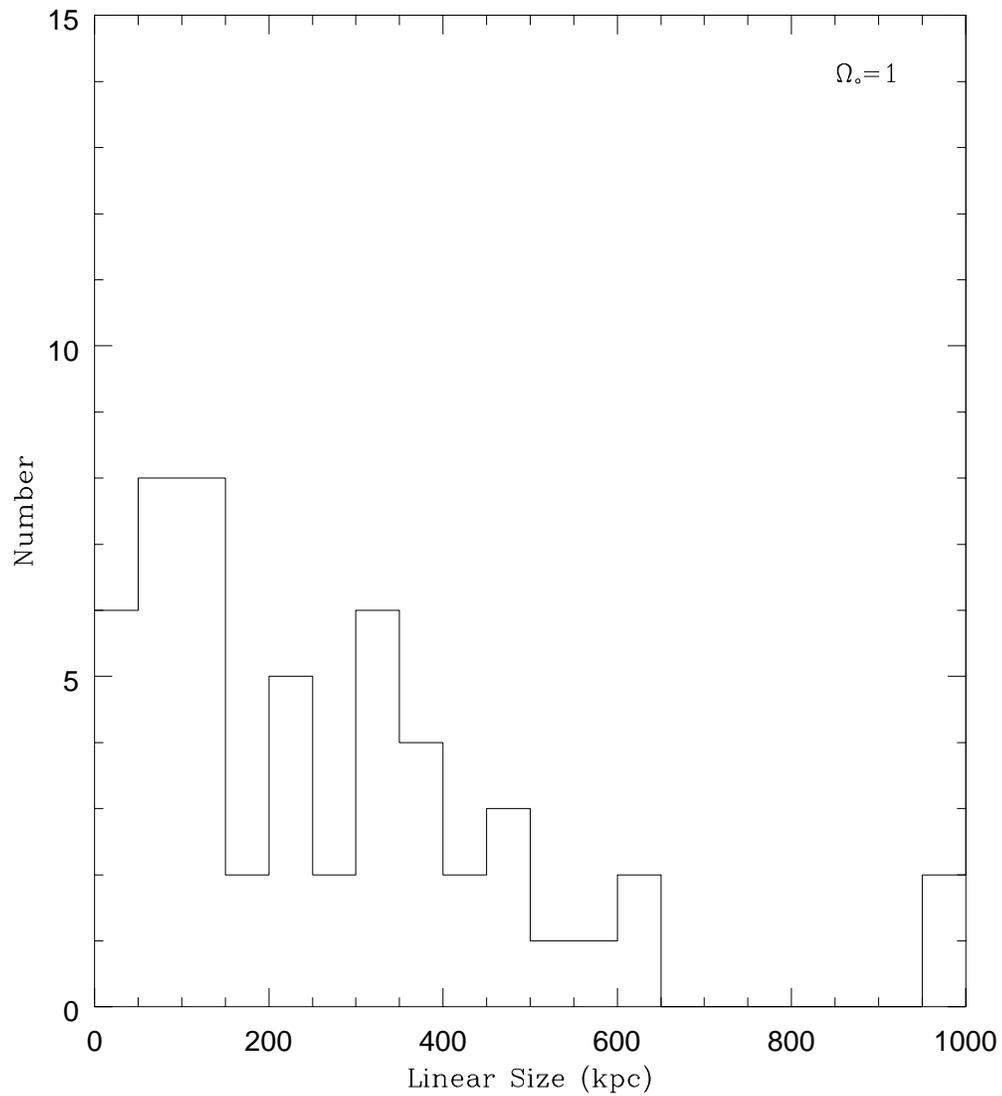

Figure 5a

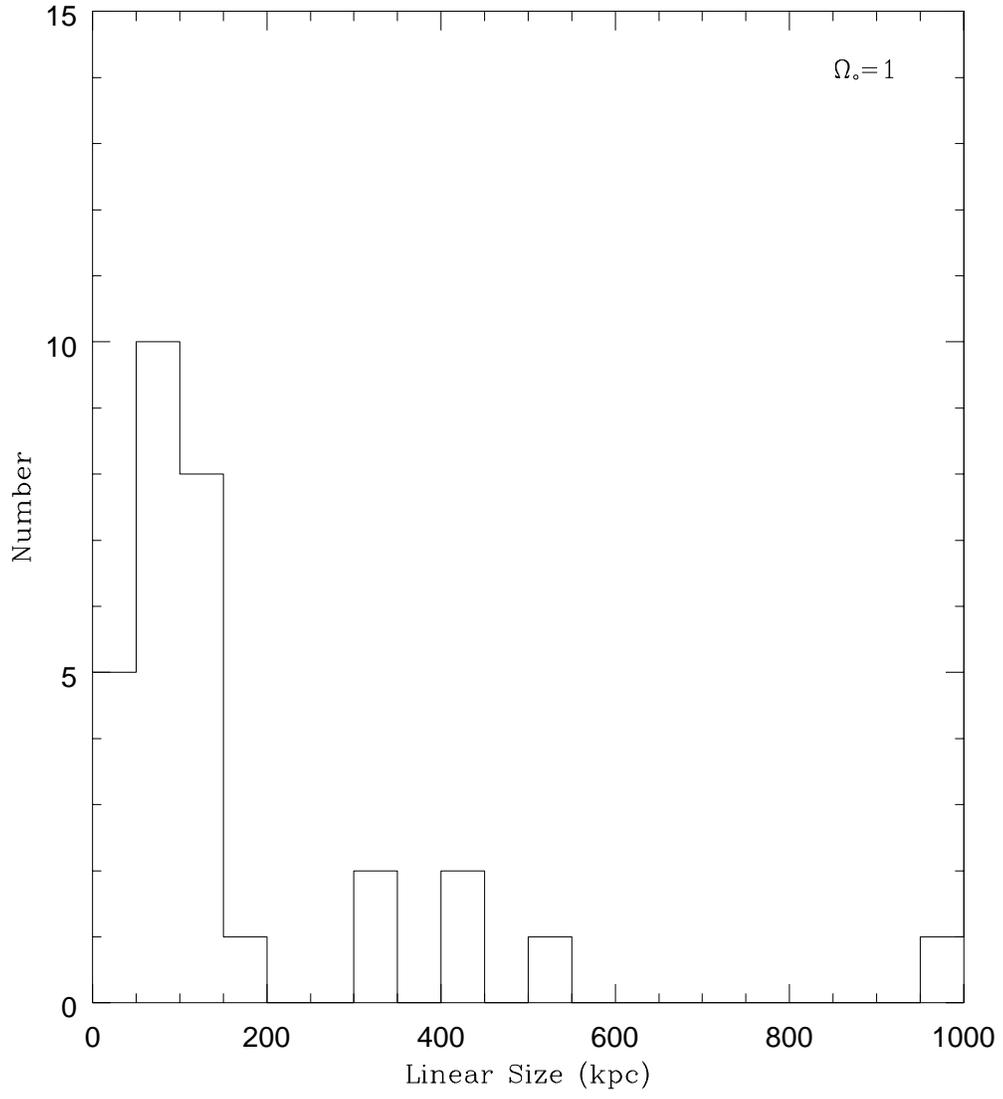

Linear Size Distribution for 6C Sample

$\Omega_o = 1$

Figure 5b